# Analysis of ZDDP content and thermal decomposition in motor oils using NAA and NMR

S. Ferguson[a], J. Johnson[a], D. Gonzales[a], C. Hobbs[b], C. Allen[a], S. Williams[a*]

[a]Department of Physics and Geosciences, Angelo State University, ASU Station #10904, San Angelo, Texas 76909, USA

[b]Department of Chemistry and Biochemistry, Angelo State University, ASU Station #10892, San Angelo, Texas 76909, USA

Zinc dialkyldithiophosphates (ZDDPs) are one of the most common anti-wear additives present in commercially-available motor oils. The ZDDP concentrations of motor oils are most commonly determined using inductively coupled plasma atomic emission spectroscopy (ICP-AES). As part of an undergraduate research project, we have determined the Zn concentrations of eight commercially-available motor oils and one oil additive using neutron activation analysis (NAA), which has potential for greater accuracy and less sensitivity to matrix effects as compared to ICP-AES. The $^{31}$P nuclear magnetic resonance ($^{31}$P-NMR) spectra were also obtained for several oil additive samples which have been heated to various temperatures in order to study the thermal decomposition of ZDDPs.

## 1. Introduction

Zinc dialkyldithiophosphates (ZDDPs) are one of the most common additives present in commercially-available motor oils. ZDDPs act as antioxidants, corrosion inhibitors, and antiwear agents (Spikes, 2004). In recent years, the ZDDP-concentrations in many motor oils have decreased substantially. For example, in 1995, when the American Petroleum Institute (API) introduced the SH service rating for motor oils to be used in gasoline engines, the maximum concentration of ZDDPs allowed was 1200 ppm (American Petroleum Institute, 2010). However, the current SN service rating, introduced in 2010, only allows a maximum ZDDP-concentration of 800 ppm (American Petroleum Institute, 2010). The decreases in ZDDP-concentrations have been due (primarily) to concerns about the effects of ZDDPs on catalytic converters. There have been studies that suggest that the presence of P in motor oils can lead to phosphate deposits on the catalysts that can reduce the conversion efficiencies of catalytic converters (Kaleli, 2001; Huang et al., 2004a). Furthermore, there have been concerns about the effects of ZDDPs on the environment. Studies have shown that when ZDDPs pollute soil and ground water they can lead to toxicities when they decompose to S and P-containing poisonous compounds (Huang et al., 2004b).

There have been concerns among some automotive enthusiasts that the lowered ZDDP-concentrations may have detrimental effects on engines. Specifically, there has been much discussion concerning the effects on engines with "flat-tappet" cam shafts and high-pressure valve springs in classic cars (which typically are not equipped with catalytic converters). The decreases in ZDDP-concentrations have been blamed for engine failures during both the "break-in" period and during normal operation. As a result, several specialty oils and additives with advertised high concentrations of ZDDPs have been developed (Smith, 2011). Due to the fact that the exact ZDDP-concentrations of many of these specialty oils and additives are not usually published and are typically considered proprietary information, there has been considerable debate among automotive enthusiasts about ZDDP-concentrations of various motor oils. It is worth noting, however, that there is some evidence that suggests that if ZDDP-concentrations exceed approximately 1800 ppm they can cause corrosion inhibitors to become less effective, resulting in an increase in corrosion-related issues (Tyger, 2013).

While there have been studies performed to determine the ZDDP-concentrations of commercially-available motor oils and additives (Smith, 2011), virtually all of them have been performed using inductively coupled plasma atomic emission spectroscopy (ICP-AES). Neutron activation analysis (NAA), however, has the potential for greater accuracy when performing trace analysis since ICP-AES is more sensitive to matrix effects and interferences (Greenberg, 2008). While NAA is occasionally used for the analysis of crude oil (Nadkarni, 2011), we are aware of only a single study in which motor oils were analyzed using NAA. Bódizs and Seif El-Nasr (1989) used NAA to determine the concentrations of trace elements in a sample of motor oil that was unused and samples that were taken from an engine crankcase after an automobile had been driven 4000 and 10000 km. Although the specific brand of motor oil that was analyzed was not mentioned, the ZDDP-concentrations of their samples were approximately 1100 ppm.

Although ZDDPs are commonly used additives in motor oils, the decomposition process is not well understood. Studies have shown that it is the decomposition products of ZDDPs, and not ZDDPs themselves, that form antiwear films (Fuller et al., 1998). While there have been numerous studies in which nuclear magnetic resonance (NMR) has been used to analyze the thermal decomposition of ZDDPs in solution (Peng et al., 1994; Fuller et al., 1998; Kapur et al., 1999), the results have sometimes been inconsistent. Furthermore, there have been no studies known to the authors of this report in which the decomposition of ZDDPs in a commercially-available oil additive has been investigated.

One of the primary goals of this study was to determine whether or not ZDDP concentrations in a complex matrix such as motor oil can be measured with greater accuracy using NAA than when using the standard method used in industry (ICP-AES). A previous report by Von Lehmden et al. (1974) compared the concentrations of trace elements in fuel oil as determined by several analytical techniques (including NAA and XRF) and concluded that "the matrix can have a pronounced effect on the reported range of concentrations" of certain elements. Secondly, we wanted to study the thermal decomposition of ZDDPs that have been heated to various temperatures using $^{31}$P-NMR and compare to the results of a previous study (Fuller et al., 1998).

## 2. Experimental

### 2.1. NAA

Eight commercially-available motor oils were chosen for analysis, as well as one oil additive. We were interested in measuring the ZDDP-concentrations of both specialty oils and more commonly used oils in order to get a better idea of what the ranges of ZDDP concentrations were in commercially-available oils. Hence, four of the oil samples analyzed had SN API service ratings, one had an SM API service rating, and four (including the additive) had no API service ratings (suggesting that the ZDDP-concentrations may exceed 800 ppm). Prior to neutron-irradiation, the samples (with masses of ~0.5 g) were sealed in polyethylene vials.

Each of the samples was irradiated using the 1.1 MW TRIGA Mark II research reactor (operating at 950 kW) at the University of Texas at Austin for 7380 s, along with two standard reference materials (NIST 1632c and IV Zn 100 ppm Lot: E2-ZN02069). The NIST 1632c reference sample, certified to have a Zn-concentration of 12.1 ± 1.3 mg/kg, was used for quality control purposes, and the second standard sample was used to determine the Zn-concentrations of the oil samples via a comparison method. During irradiation, the thermal neutron flux density was ~2 x $10^{12}$ cm$^{-2}$ s$^{-1}$.

The ZDDP-concentrations of the oil samples were determined by analyzing the 1.115 MeV $^{65}$Zn photopeaks in the radiation spectra. $^{65}$Zn is a relatively long-lived isotope with a half-life of approximately 244 days (Landsberger, 1994). Radiation from the samples was detected using a high-purity Ge detector with a thin Be window for durations of 10800 s. A typical spectrum is shown in Fig. 1.

### 2.2. NMR

Five samples of Comp Cams Engine Break-In Oil Additive were used for NMR analysis. Sample A was not heated. Sample B was heated at 100 °C (which is slightly higher than the operating temperature of most gasoline engines) for 48 h. Samples C and D were heated at 150 °C for 6 h and 24 h, respectively. Sample E was heated at 200 °C for 1 h. After the samples had been allowed to cool, the $^{31}$P-NMR spectra were obtained using an Agilent 400 spectrometer operating at 161.839 MHz.

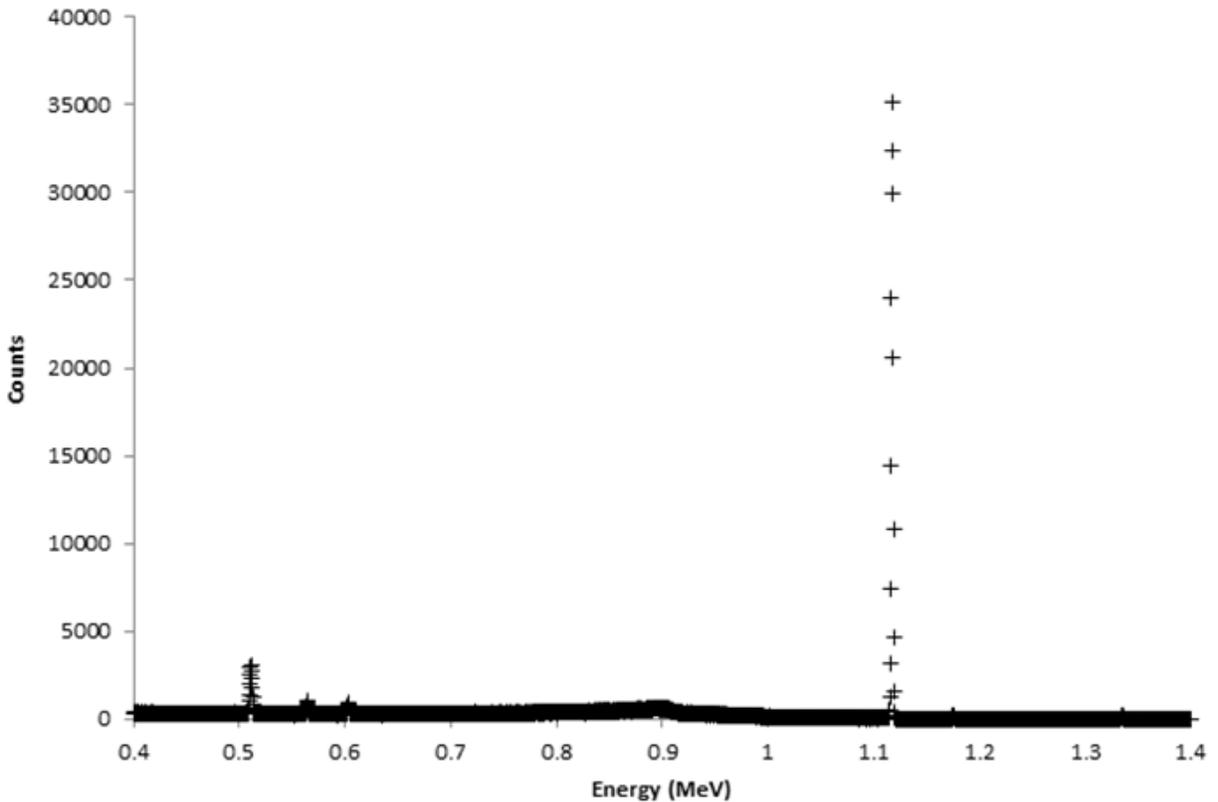

Fig. 1. A typical NAA spectrum. The prominent photopeak associated with $^{65}$Zn can be seen at 1.115 MeV.

### 3. Results and Discussion

*3.1. NAA Results*

The Zn-concentrations of the samples are presented in Table 1. It was assumed that the Zn-contents of the oils samples were entirely in the form of ZDDPs and that the oils contained no other additives with Zn. The data provided by one manufacturer indicated that this assumption was correct (Tyger, 2013) and the authors of this report know of no common oil additives, other than ZDDPs, that contain Zn.

As might be expected, the specialty oils with no API service ratings all had ZDDP concentrations that exceeded API SN service rating specifications. Two of the four oils with SN service ratings also had ZDDP concentrations that exceeded the specifications (although only slightly). The sole oil with the API SM service rating (which is formulated for diesel engines, but commonly used in many engines equipped with flat-tappet cams due to rumors of high ZDDP-concentrations) had a ZDDP-content comparable to the specialty oils analyzed. The lone additive tested (Comp Cams Engine Break-In Oil Additive) had a ZDDP-concentration that was more than five times greater than the concentrations of any of the oils analyzed.

| Brand | API Rating | Zn (ppm) | Uncertainty (ppm) | Detection Limits (ppm) |
|---|---|---|---|---|
| Red Line 10W30 | N/A | 1113 | 12.75 | 0.68 |
| Super Tech 10W30 | SN | 675.44 | 7.98 | 0.82 |
| Valvoline Conventional 10W30 | SN | 756.24 | 8.86 | 0.78 |
| Valvoline VR1 10W30 | N/A | 1296.78 | 14.83 | 0.86 |
| Comp Cams Engine Break-In Oil Additive | N/A | 9548.28 | 106.97 | 2.95 |
| Brad Penn Penn-Grade 1 10W30 | N/A | 1630.61 | 18.6 | 1.09 |
| Royal Purple 10W30 | SN | 842.03 | 9.75 | 0.67 |
| Shell Rotella T5 10W30 | SM | 1242.12 | 14.41 | 1.13 |
| Lucas 10W30 | SN | 824.45 | 9.58 | 0.87 |
| NIST 1632c | N/A | 15.8 | 0.69 | 1.69 |
| IV Zn 100 ppm Lot: E2-ZN02069 | N/A | 99.15 | 1.49 | 0.35 |

Table 1. ZDDP-concentrations of the samples as determined via NAA.

Total uncertainties were calculated using software which accounts for experimental uncertainties resulting from factors such as detector "dead time," as well as statistical uncertainty. The uncertainties associated with the Zn-content of the motor oils analyzed via NAA are all less than 1.2%. This is significantly better than the accuracy typically achieved when analyzing oils samples using ICP-AES. A recent study (Fox, 2005) analyzing Zn-content in lubricating grease via ICP-AES achieved uncertainties ranging from 4.8% to 5.8%, despite the fact that the samples contained much higher ZDDP concentrations (10.6 wt% Zn) than those used in the present study.

*3.2. NMR Results*

The $^{31}$P-NMR spectra of each of the five samples can be seen in Fig. 2. Commercial ZDDPs are mixtures of both basic and neutral ZDDPs (Fuller et al., 1998), and the signals for ZDDPs in neutral form are found at lower ppm than those in basic form. Furthermore, the ZDDPs used in most motor oils consist of both secondary (C-4) and primary (C-8) alkyl groups. Thus, we expected to see three signals from both the neutral and basic ZDDPs in the $^{31}$P-NMR spectra: one corresponding only to secondary alkyl groups, one corresponding only to primary alkyl groups, and one corresponding to secondary and primary alkyl groups. The results shown in Fig. 2 suggest that the magnetic field used to obtain our $^{31}$P-NMR spectra was not strong enough for all six of the expected signals to be resolved. Rather, we were only able to resolve the signals from the basic and neutral ZDDPs.

The data shown in Fig. 2 suggests that only ZDDPs in Samples C and D, which were heated at 150 °C for 6 and 24 h, respectively, show any signs of thermal decomposition. In the case of Sample D, the thermal decomposition was nearly complete. The results of a previous study performed by Fuller et al. (1998), in which the thermal decomposition of an oil solution consisting of 1.2 wt% ZDDP was investigated, were, for the most part, similar. However, there are a few significant differences between the results presented here and those of Fuller et al. (1998). While the $^{31}$P-NMR spectrum of Sample E, which was heated at 200 °C for 1 h, suggests no significant thermal decomposition, Fuller et al. (1998) found that neither "ZDDP, nor any oil soluble ZDDP decomposition intermediates and/or products [were] left in solution after 1 h of heating at 200 °C." In the present study, Sample E was placed in an oven at 200 °C for 1 h, thus the sample itself was not at a temperature of 200 °C for the entire hour. It is not known if Fuller et al. (1998) followed a similar procedure, or if the sample in their study was held at 200 °C for an entire hour. While this is one possible explanation of the discrepancy in the results, the relatively good signal-to-noise ratio of the spectrum of Sample E in Fig. 2, suggests that no significant thermal decomposition occurred when our sample was heated in an oven at 200 °C for 1 h. Furthermore, although no chemical shifts can be seen in the data shown in Fig. 2, the results of Fuller et al. (1998) found slight chemical shifts to higher ppm as samples were heated at higher temperatures and for longer durations. It is possible that this discrepancy is the result of differences in the nature of the alkyl groups, solvents, and concentrations of the samples used in the studies, which have been shown to affect $^{31}$P chemical shifts (Harrison and Kikabhai, 1987).

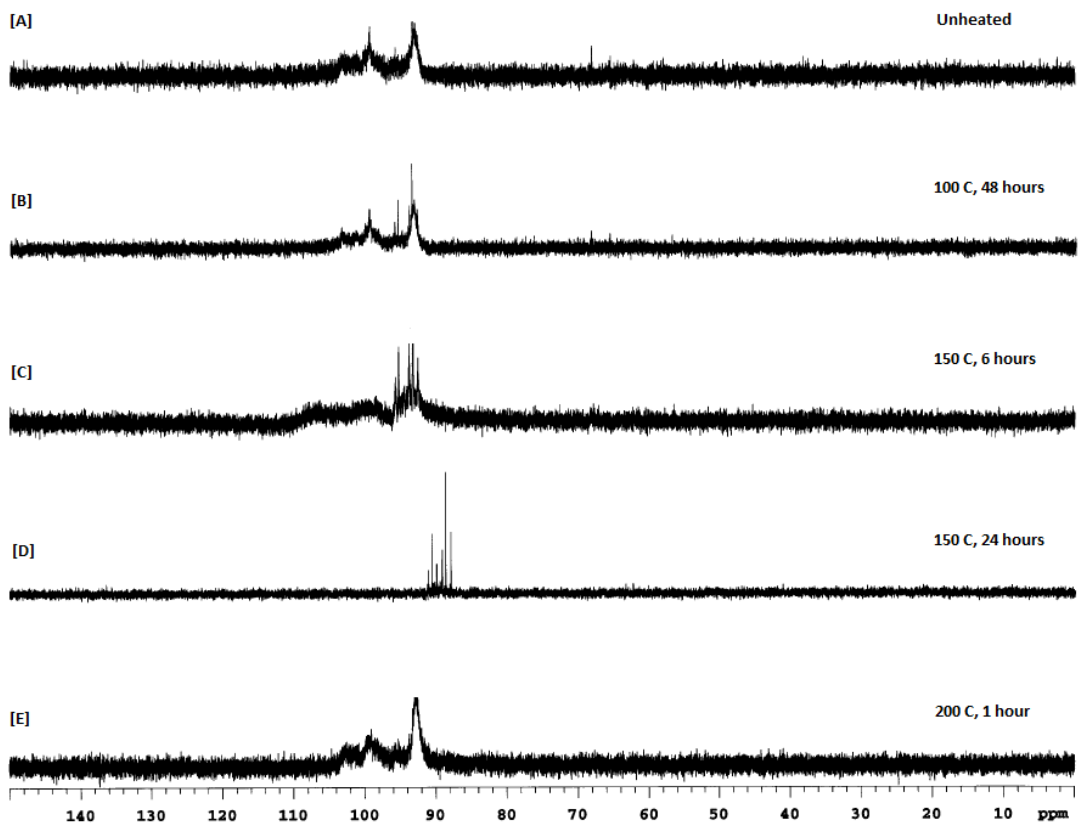

Fig. 2. $^{31}$P-NMR spectra of five samples of the oil additive, heated to different temperatures for various durations.

## 4. Conclusions

We have determined the Zn concentrations of eight commercially-available motor oils and one oil additive as part of an undergraduate research project. The results suggest that NAA is a suitable method for determining the concentrations of trace elements in oils. While the facilities needed for NAA (nuclear reactors, etc.) are not always easily accessible, it seems as though the potential for greater accuracy and less sensitivity to matrix effects as compared to ICP-AES (Greenberg, 2008) might encourage motor oil manufacturers to consider using NAA for quality control purposes.

The $^{31}$P-NMR spectra obtained for five samples of an oil additive containing relatively high concentrations of ZDDP heated at various temperatures for various durations of time suggest that thermal decomposition occurs at temperatures as low as 150 °C. No ZDDP decomposition was observed for the sample heated at 200 °C for 1 h, in disagreement with a report by Fuller et al. (1998). Differences in the results presented here and the results obtained by Fuller et al. (1998) may due to differences in the way in which the samples were prepared or differences in the nature of the alkyl groups, solvents, and concentrations. In the future, it would be interesting to study the $^{31}$P-NMR spectra of samples heated at 150 °C for durations of time between 6 h and 24 h, in order to gain information about the time dependence of the thermal degradation of ZDDPs at this temperature.

## Acknowledgements

The authors thank Sheldon Landsberger and Tracy Tipping at the University of Texas for assistance with NAA. The authors also thank the Angelo State University Center for Innovation in Teaching and Research for financial support.